# Multi-Erasure Locally Recoverable Codes Over Small Fields For Flash Memory Array

**Pengfei Huang**\*, **Eitan Yaakobi**†, and **Paul H. Siegel**\*

\*Electrical and Computer Engineering Dept., University of California, San Diego, La Jolla, CA 92093 U.S.A
†Computer Science Dept., Technion – Israel Institute of Technology, Haifa 32000, Israel
{*pehuang,psiegel*}*@ucsd.edu*, *yaakobi@cs.technion.ac.il*

*Abstract*—Erasure codes play an important role in storage systems to prevent data loss. In this work, we study a class of erasure codes called Multi-Erasure Locally Recoverable Codes (ME-LRCs) for flash memory array. Compared to previous related works, we focus on the construction of ME-LRCs over small fields. We first develop upper and lower bounds on the minimum distance of ME-LRCs. These bounds explicitly take the field size into account. Our main contribution is to propose a general construction of ME-LRCs based on generalized tensor product codes, and study their erasure-correcting property. A decoding algorithm tailored for erasure recovery is given. We then prove that our construction yields optimal ME-LRCs with a wide range of code parameters. Finally, we present several families of ME-LRCs over different fields.

## I. INTRODUCTION

Recently, erasure codes with both local and global erasure-correcting properties have received considerable interests [5], [7], [14]–[16], [19], thanks to their promising application in storage systems. The idea behind is when a few erasures occur, these erasures can be corrected fast using only local parities. If the number of erasures exceeds the local erasure-correcting capability, then the global parities are invoked.

In this paper, we consider this kind of erasure codes with both local and global erasure-correcting capabilities for a $\rho \times n_0$ flash memory array [3], [5], where each row contains some local parities, and additional global parities are distributed in the array. More specifically, let us give the formal definition of this class of erasure codes as follows.

**Definition 1.** *Consider a code $\mathcal{C}$ over a finite field $\mathbb{F}_q$ consisting of $\rho \times n_0$ arrays such that:*

1) *Each row in each array in $\mathcal{C}$ belongs to a linear local code $\mathcal{C}_0$ with length $n_0$ and minimum distance $d_0$ over $\mathbb{F}_q$.*
2) *Reading the symbols of $\mathcal{C}$ row-wise, $\mathcal{C}$ is a linear code with length $\rho n_0$, dimension $k$, and minimum distance $d$ over $\mathbb{F}_q$.*

*Then, we say that $\mathcal{C}$ is a $(\rho, n_0, k; d_0, d)_q$ **Multi-Erasure Locally Recoverable Code (ME-LRC)**.*

Thus, a $(\rho, n_0, k; d_0, d)_q$ ME-LRC can locally correct $d_0 - 1$ erasures in each row, and is guaranteed to correct a total of $d - 1$ erasures anywhere in the array.

Our work is motivated by a recent work by Blaum and Hetzler [5]. In this work, the authors studied ME-LRCs where each row is a maximum distance separable (MDS) code, and gave code constructions with field size $q \geqslant \max\{\rho, n_0\}$ from integrated interleaving codes [9]. As in Definition 1, we generalize the definition of the codes studied in [5] such that each row is not necessary an MDS code. Moreover, we give upper and lower bounds on the minimum distance of ME-LRCs.

For the code construction, we focus on constructing ME-LRCs over small fields. We propose a general construction method from generalized tensor product codes [13], [22]. In contrast to [5], our construction is more flexible in the sense that we do not require field size $q \geqslant \max\{\rho, n_0\}$, and it can even generate binary ME-LRCs. For $2d_0 \geqslant d$, we will show that our construction gives *optimal* ME-LRCs over $\mathbb{F}_q$, with respect to our upper bound on the minimum distance.

There are other related works. If the local code $\mathcal{C}_0$ of an ME-LRC is a single parity-check code, then the ME-LRC is a Locally Repairable Code (LRC) [7], [14], [15], [19], which is a class of codes where a single erased symbol can be recovered by accessing $r$ other symbols, and here $r$ is called the locality of this code. LRCs over small fields were studied in [6], [8], [10]–[12], [20], [24]. In [16], $(r, \delta)$ LRCs were proposed. A code $\mathcal{C}$ is called an $(r, \delta)$ LRC if for every coordinate, there exists a punctured code (i.e., a repair set) of $\mathcal{C}$ with support containing this coordinate, whose length is at most $r + \delta - 1$, and whose minimum distance is at least $\delta$. ME-LRCs studied in this paper can be seen as $(r, \delta)$ LRCs with disjoint repair sets. However, the previous existing constructions of $(r, \delta)$ LRCs require relatively large field size $q$, which is at least as large as the code length [16], [18], [19]. A very recent work [1] gives explicit constructions of $(r, \delta)$ LRCs with disjoint repair sets over field $\mathbb{F}_q$ from algebraic curves, whose repair sets have size $r + \delta - 1 = \sqrt{q}$ or $r + \delta - 1 = \sqrt{q} + 1$. In contrast, our construction is more flexible, and can generate codes with field size $q$ even smaller than the repair set size. In addition, a Partial MDS (PMDS) code [4] can also be seen as an ME-LRC with certain constraints.

The goal of this paper is to study ME-LRCs over small fields. The remainder of the paper is organized as follows. In Section II, we study field size dependent upper and lower bounds for ME-LRCs. In Section III, we propose a general construction of ME-LRCs. The erasure-correcting property of these codes is studied and an erasure decoding algorithm is also presented. In Section IV, we give several families of ME-LRCs over different fields. Section V concludes the paper.

Throughout the paper, we use the notation $[n]$ to define the set $\{1, \ldots, n\}$. For a length-$n$ vector $v$ over $\mathbb{F}_q$ and a set $\mathcal{I} \subseteq [n]$, the vector $v_\mathcal{I}$ denotes the restriction of the vector $v$ to coordinates in the set $\mathcal{I}$, and $w_q(v)$ represents the Hamming weight of the vector over $\mathbb{F}_q$. The transpose of a matrix $H$ is written as $H^T$. A linear code over $\mathbb{F}_q$ of length $n$, dimension $k$, and minimum distance $d$ will be denoted by $[n, k, d]_q$, whose redundancy is $n - k$. For a code with only one codeword, its minimum distance is defined as $\infty$.

## II. UPPER AND LOWER BOUNDS FOR ME-LRCs

In this section, we study field size dependent upper and lower bounds on the minimum distance of ME-LRCs. The upper bound derived here will be used to prove the optimality of our construction for ME-LRCs in the following sections.

Now, we give an upper bound on the minimum distance of a $(\rho, n_0, k; d_0, d)_q$ ME-LRC. The shortening technique employed here was previously used for bounds for low-density parity-check (LDPC) codes [2] and locally repairable codes [6].

Let $d_{opt}^{(q)}[n, k]$ denote the largest possible minimum distance of a linear code of length $n$ and dimension $k$ over $\mathbb{F}_q$, and let $k_{opt}^{(q)}[n, d]$ denote the largest possible dimension of a linear code of length $n$ and minimum distance $d$ over $\mathbb{F}_q$.

**Lemma 2.** *For any $(\rho, n_0, k; d_0, d)_q$ ME-LRC $\mathcal{C}$, the minimum distance $d$ satisfies*

$$d \leqslant \min_{0 \leqslant x \leqslant \lceil \frac{k}{k^*} \rceil - 1} \left\{ d_{opt}^{(q)}[\rho n_0 - x n_0, k - x k^*] \right\}, \quad (1)$$

*and the dimension satisfies*

$$k \leqslant \min_{0 \leqslant x \leqslant \lceil \frac{k}{k^*} \rceil - 1} \left\{ x k^* + k_{opt}^{(q)}[\rho n_0 - x n_0, d] \right\}, \quad (2)$$

*where $k^* = k_{opt}^{(q)}[n_0, d_0]$.*

*Proof:* For the case of $x = 0$, it is trivial. For $1 \leqslant x \leqslant \lceil \frac{k}{k^*} \rceil - 1$, $x \in \mathbb{Z}^+$, let $\mathcal{I}$ represent the set of the coordinates of the first $x$ rows in the array. Thus, $|\mathcal{I}| = xn_0$. First, consider the code $\mathcal{C}_\mathcal{I} = \{c_\mathcal{I} : c \in \mathcal{C}\}$ whose dimension is denoted by $k_\mathcal{I}$, which satisfies $k_\mathcal{I} \leqslant xk^*$. Then, we consider the code $\mathcal{C}_\mathcal{I}^0 = \{c_{[\rho n_0]\setminus \mathcal{I}} : c_\mathcal{I} = \mathbf{0} \text{ and } c \in \mathcal{C}\}$. Since the code $\mathcal{C}$ is linear, the size of the code $\mathcal{C}_\mathcal{I}^0$ is $q^{k-k_\mathcal{I}}$ and it is a linear code as well. Moreover, the minimum distance $\hat{d}$ of the code $\mathcal{C}_\mathcal{I}^0$ is at least $d$, i.e., $\hat{d} \geqslant d$.

Thus, we get an upper bound on the minimum distance $d$,

$$d \leqslant \hat{d} \leqslant d_{opt}^{(q)}[\rho n_0 - |\mathcal{I}|, k - k_\mathcal{I}]$$
$$\leqslant d_{opt}^{(q)}[\rho n_0 - x n_0, k - x k^*].$$

Similarly, we also get an upper bound on the dimension $k$,

$$k - k_\mathcal{I} \leqslant k_{opt}^{(q)}[\rho n_0 - |\mathcal{I}|, \hat{d}] \leqslant k_{opt}^{(q)}[\rho n_0 - x n_0, d].$$

Therefore, we conclude that

$$k \leqslant k_{opt}^{(q)}[\rho n_0 - x n_0, d] + k_\mathcal{I} \leqslant k_{opt}^{(q)}[\rho n_0 - x n_0, d] + x k^*.$$

∎

An asymptotic lower bound for ME-LRCs with local MDS codes was given in [1]. Here, by simply adapting the Gilbert-Varshamov (GV) bound [17], we have the following GV-like lower bound on ME-LRCs of finite length without specifying local codes.

**Lemma 3.** *A $(\rho, n_0, k; d_0, d)_q$ ME-LRC $\mathcal{C}$ exists, if*

$$\sum_{i=0}^{d-2} \binom{\rho(n_0 - \lceil \log_q(\sum_{j=0}^{d_0-2} \binom{n_0-1}{j}(q-1)^j) \rceil) - 1}{i}(q-1)^i$$
$$< q^{\rho(n_0 - \lceil \log_q(\sum_{j=0}^{d_0-2} \binom{n_0-1}{j}(q-1)^j) \rceil) - k}. \quad (3)$$

*Proof:* See Appendix A. ∎

## III. ME-LRCs FROM GENERALIZED TENSOR PRODUCT CODES: CONSTRUCTION AND DECODING

Tensor product codes, first proposed by Wolf in [22], are a family of binary error-correcting codes defined by a parity-check matrix that is the tensor product of the parity-check matrices of two constituent codes. Later, they were generalized in [13]. In this section, we first introduce generalized tensor product codes over $\mathbb{F}_q$. Then, we give a general construction of ME-LRCs from generalized tensor product codes. The minimum distance of the constructed ME-LRCs is determined, and a corresponding decoding algorithm tailored for erasure correction is proposed.

### A. Generalized Tensor Product Codes over $\mathbb{F}_q$

We start by presenting the tensor product operation of two matrices $H'$ and $H''$. Let $H'$ be the parity-check matrix of a code with length $n'$ and dimension $n' - v$ over $\mathbb{F}_q$. The matrix $H'$ can be considered as a $v$ (row) by $n'$ (column) matrix over $\mathbb{F}_q$ or as a 1 (row) by $n'$ (column) matrix of elements from $\mathbb{F}_{q^v}$. Let $H'$ be the vector $H' = [h_1'\ h_2'\ \cdots\ h_{n'}']$, where $h_j'$, $1 \leqslant j \leqslant n'$, are elements of $\mathbb{F}_{q^v}$. Let $H''$ be the parity-check matrix of a code of length $\ell$ and dimension $\ell - \lambda$ over $\mathbb{F}_{q^v}$. We denote $H''$ by

$$H'' = \begin{bmatrix} h_{11}'' & \cdots & h_{1\ell}'' \\ \vdots & \ddots & \vdots \\ h_{\lambda 1}'' & \cdots & h_{\lambda \ell}'' \end{bmatrix},$$

where $h_{ij}''$, $1 \leqslant i \leqslant \lambda$ and $1 \leqslant j \leqslant \ell$, are elements of $\mathbb{F}_{q^v}$.

The tensor product of the matrices $H'$ and $H''$ is defined as

$$H_{TP} = H'' \bigotimes H' = \begin{bmatrix} h_{11}''H' & \cdots & h_{1\ell}''H' \\ \vdots & \ddots & \vdots \\ h_{\lambda 1}''H' & \cdots & h_{\lambda \ell}''H' \end{bmatrix},$$

where $h_{ij}''H' = [h_{ij}''h_1'\ h_{ij}''h_2'\ \cdots\ h_{ij}''h_{n'}']$, $1 \leqslant i \leqslant \lambda$ and $1 \leqslant j \leqslant \ell$, and the products of elements are calculated according to the rules of multiplication for elements over $\mathbb{F}_{q^v}$. The matrix $H_{TP}$ is considered as a $v\lambda \times n'\ell$ matrix of elements from $\mathbb{F}_q$.

Our construction of ME-LRCs is based on generalized tensor product codes [13]. Define the matrices $H_i'$ and $H_i''$ for $i = 1, 2, \ldots, \mu$ as follows. The matrix $H_i'$ is a $v_i \times n'$ matrix over $\mathbb{F}_q$ such that the $(v_1 + v_2 + \cdots + v_i) \times n'$ matrix

$$B_i = \begin{bmatrix} H_1' \\ H_2' \\ \vdots \\ H_i' \end{bmatrix}$$

is a parity-check matrix of an $[n', n' - v_1 - v_2 - \cdots - v_i, d_i']_q$ code, where $d_1' < d_2' < \cdots < d_i'$. The matrix $H_i''$ is a $\lambda_i \times \ell$ matrix over $\mathbb{F}_{q^{v_i}}$, which is a parity-check matrix of an $[\ell, \ell - \lambda_i, \delta_i]_{q^{v_i}}$ code.

We define a $\mu$-level generalized tensor product code over $\mathbb{F}_q$ as a linear code having a parity-check matrix in the form of the following $\mu$-level tensor product structure

$$H = \begin{bmatrix} H_1'' \otimes H_1' \\ H_2'' \otimes H_2' \\ \vdots \\ H_\mu'' \otimes H_\mu' \end{bmatrix}. \quad (4)$$

We denote this code by $\mathcal{C}_{GTP}^\mu$. Its length is $n_t = n'\ell$ and the dimension is $k_t = n_t - \sum_{i=1}^\mu v_i \lambda_i$.

Let us give an example of a 2-level generalized tensor product code $\mathcal{C}_{GTP}^2$ over $\mathbb{F}_2$.

**Example 1.** Let $H_1' = [1\ 1\ 1\ 1\ 1\ 1\ 1]$ over $\mathbb{F}_2$, and

$$H_2' = \begin{bmatrix} 0 & 0 & 0 & 1 & 1 & 1 & 1 \\ 0 & 1 & 1 & 0 & 0 & 1 & 1 \\ 1 & 0 & 1 & 0 & 1 & 0 & 1 \end{bmatrix}$$

over $\mathbb{F}_2$. Let $H_1''$ be a $3 \times 3$ identity matrix

$$H_1'' = \begin{bmatrix} 1 & 0 & 0 \\ 0 & 1 & 0 \\ 0 & 0 & 1 \end{bmatrix}$$

over $\mathbb{F}_2$, and $H_2'' = [1\ 1\ 1]$ over $\mathbb{F}_8$. Hence, in this construction, we have the following parameters: $n' = 7$, $\ell = 3$, $v_1 = 1$, $v_2 = 3$, $\lambda_1 = 3$, $\lambda_2 = 1$, $d_1' = 2$, $d_2' = 4$, $\delta_1 = \infty$, and $\delta_2 = 2$. The binary parity-check matrix $H$ of the 2-level tensor product code $\mathcal{C}_{GTP}^2$ is

$$H = \begin{bmatrix} H_1'' \otimes H_1' \\ H_2'' \otimes H_2' \end{bmatrix}$$

$$= \begin{bmatrix} 1\ 1\ 1\ 1\ 1\ 1\ 1 & 0\ 0\ 0\ 0\ 0\ 0\ 0 & 0\ 0\ 0\ 0\ 0\ 0\ 0 \\ 0\ 0\ 0\ 0\ 0\ 0\ 0 & 1\ 1\ 1\ 1\ 1\ 1\ 1 & 0\ 0\ 0\ 0\ 0\ 0\ 0 \\ 0\ 0\ 0\ 0\ 0\ 0\ 0 & 0\ 0\ 0\ 0\ 0\ 0\ 0 & 1\ 1\ 1\ 1\ 1\ 1\ 1 \\ 0\ 0\ 0\ 1\ 1\ 1\ 1 & 0\ 0\ 0\ 1\ 1\ 1\ 1 & 0\ 0\ 0\ 1\ 1\ 1\ 1 \\ 0\ 1\ 1\ 0\ 0\ 1\ 1 & 0\ 1\ 1\ 0\ 0\ 1\ 1 & 0\ 1\ 1\ 0\ 0\ 1\ 1 \\ 1\ 0\ 1\ 0\ 1\ 0\ 1 & 1\ 0\ 1\ 0\ 1\ 0\ 1 & 1\ 0\ 1\ 0\ 1\ 0\ 1 \end{bmatrix}.$$

The code length is $n_t = n'\ell = 21$ and the dimension is $k_t = n_t - \sum_{i=1}^2 v_i \lambda_i = 15$. The minimum distance is $d_t = 4$. □

By adapting Theorem 2 in [13] from the field $\mathbb{F}_2$ to $\mathbb{F}_q$, we directly have the following theorem on the minimum distance of $\mathcal{C}_{GTP}^\mu$ over $\mathbb{F}_q$.

**Theorem 4.** *The minimum distance $d_t$ of a generalized tensor product code $\mathcal{C}_{GTP}^\mu$ over $\mathbb{F}_q$ satisfies*

$$d_t \geq \min\{\delta_1, \delta_2 d_1', \delta_3 d_2', \ldots, \delta_\mu d_{\mu-1}', d_\mu'\}.$$

*Proof:* See Appendix B. ∎

### B. Construction of ME-LRCs

Now, we present a general construction of ME-LRCs based on generalized tensor product codes.

**Construction A**

**Step 1:** Choose $v_i \times n'$ matrices $H_i'$ over $\mathbb{F}_q$ and $\lambda_i \times \ell$ matrices $H_i''$ over $\mathbb{F}_{q^{v_i}}$, for $i = 1, 2, \ldots, \mu$, which satisfy the following two properties:

1) The parity-check matrix $H_1'' = \mathbf{I}_{\ell \times \ell}$, i.e., an $\ell \times \ell$ identity matrix.
2) The matrices $H_i'$ (or $B_i$), $1 \leq i \leq \mu$, and $H_j''$, $2 \leq j \leq \mu$, are chosen such that $d_\mu' \leq \delta_j d_{j-1}'$, for $j = 2, 3, \cdots, \mu$.

**Step 2:** Generate a parity-check matrix $H$ according to (4) with the matrices $H_i'$ and $H_i''$, for $i = 1, 2, \ldots, \mu$. The constructed code corresponding to the parity-check matrix $H$ is referred to as $\mathcal{C}_A$. ∎

**Theorem 5.** *The code $\mathcal{C}_A$ is a $(\rho, n_0, k; d_0, d)_q$ ME-LRC with parameters $\rho = \ell$, $n_0 = n'$, $k = n'\ell - \sum_{i=1}^\mu v_i \lambda_i$, $d_0 = d_1'$, and $d = d_\mu'$.*

*Proof:* According to Construction A, the code parameters $\rho$, $n_0$, $k$, and $d_0$ can be easily determined. In the following, we prove that $d = d_\mu'$.

Since $\delta_1 = \infty$ ($H_1''$ is the identity matrix) and $d_\mu' \leq \delta_i d_{i-1}'$ for all $i = 2, 3, \ldots, \mu$, from Theorem 4, $d \geq d_\mu'$.

Now, we show that $d \leq d_\mu'$. For $i = 1, 2, \ldots, \mu$, let $H_i' = [h_1'(i), \ldots, h_{n'}'(i)]$ over $\mathbb{F}_{q^{v_i}}$, and let $[h_{11}''(i), \ldots, h_{\lambda_i 1}''(i)]^T$ over $\mathbb{F}_{q^{v_i}}$ be the first column of $H_i''$. Since the code with parity-check matrix $B_\mu$ has minimum distance $d_\mu'$, there exist $d_\mu'$ columns of $B_\mu$, say in the set of positions $J = \{b_1, b_2, \ldots, b_{d_\mu'}\}$, which are linearly dependent; that is, $\sum_{j \in J} \alpha_j h_j'(i) = 0$, $\alpha_j \in \mathbb{F}_q$, for $i = 1, 2, \ldots, \mu$. Thus, we have $\sum_{j \in J} \alpha_j h_{p1}''(i) h_j'(i) = h_{p1}''(i) \left( \sum_{j \in J} \alpha_j h_j'(i) \right) = 0$, for $p = 1, 2, \ldots, \lambda_i$ and $i = 1, 2, \ldots, \mu$. That is, the columns in positions $b_1, b_2, \ldots, b_{d_\mu'}$ of $H$ are linearly dependent. ∎

### C. Erasure Decoding of ME-LRCs

We present a decoding algorithm for the ME-LRC $\mathcal{C}_A$ from Construction A, tailored for erasure correction. The decoding algorithm for error correction for generalized tensor product codes can be found in [13].

Let the symbol ? represent an erasure and "e" denote a decoding failure. The erasure decoder $\mathcal{D}_A : (\mathbb{F}_q \cup \{?\})^{n'\ell} \to \mathcal{C}_A \cup \{\text{"e"}\}$ for an ME-LRC $\mathcal{C}_A$ consists of two kinds of component decoders $\mathcal{D}_i'$ and $\mathcal{D}_i''$ for $i = 1, 2, \ldots, \mu$ described below.

**a)** First, the decoder for a coset of the code $\mathcal{C}_i'$ with parity-check matrix $B_i$, $i = 1, 2, \ldots, \mu$, is denoted by

$$\mathcal{D}_i' : (\mathbb{F}_q \cup \{?\})^{n'} \times (\mathbb{F}_q \cup \{?\})^{\sum_{j=1}^i v_j} \to (\mathbb{F}_q \cup \{?\})^{n'}$$

which uses the following decoding rules: for a length-$n'$ input vector $\mathbf{y}'$, and a length-$\sum_{j=1}^i v_j$ syndrome vector $\mathbf{s}'$ without erasures, $\mathcal{D}_i'(\mathbf{y}', \mathbf{s}') = \mathbf{c}'$, if $\mathbf{y}'$ agrees with exactly one codeword $\mathbf{c}' \in \mathcal{C}_i' + \mathbf{e}$ on the entries with values in $\mathbb{F}_q$, where the vector $\mathbf{e}$ is a coset leader determined by both the code $\mathcal{C}_i'$ and the syndrome vector $\mathbf{s}'$; otherwise, $\mathcal{D}_i'(\mathbf{y}', \mathbf{s}') = \mathbf{y}'$. Therefore, if the length-$n'$ input vector $\mathbf{y}'$ is a codeword in $\mathcal{C}_i' + \mathbf{e}$ with $d_i' - 1$ or less erasures and the syndrome vector $\mathbf{s}'$ is not erased, then the decoder $\mathcal{D}_i'$ can return the correct codeword.

**b)** Second, the decoder for the code $\mathcal{C}_i''$ with parity-check matrix $H_i''$, $i = 1, 2, \ldots, \mu$, is denoted by

$$\mathcal{D}_i'' : (\mathbb{F}_{q^{v_i}} \cup \{?\})^\ell \to (\mathbb{F}_{q^{v_i}} \cup \{?\})^\ell$$

which uses the following decoding rules: for a length-$\ell$ input vector $\boldsymbol{y}''$, $\mathcal{D}_i''(\boldsymbol{y}'') = \boldsymbol{c}''$, if $\boldsymbol{y}''$ agrees with exactly one codeword $\boldsymbol{c}'' \in \mathcal{C}_i''$ on the entries with values in $\mathbb{F}_{q^{v_i}}$; otherwise, $\mathcal{D}_i''(\boldsymbol{y}'') = \boldsymbol{y}''$. Therefore, if the length-$\ell$ input vector $\boldsymbol{y}''$ is a codeword in $\mathcal{C}_i''$ with $\delta_i - 1$ or less erasures, then the decoder $\mathcal{D}_i''$ can successfully return the correct codeword.

The erasure decoder $\mathcal{D}_A$ for the code $\mathcal{C}_A$ is summarized in Algorithm 1 below. Let the input word of length $n'\ell$ for the decoder $\mathcal{D}_A$ be $\boldsymbol{y} = (\boldsymbol{y}_1, \boldsymbol{y}_2, \ldots, \boldsymbol{y}_\ell)$, where each component $\boldsymbol{y}_i \in (\mathbb{F}_q \cup \{?\})^{n'}$, $i = 1, \ldots, \ell$. The vector $\boldsymbol{y}$ is an erased version of a codeword $\boldsymbol{c} = (\boldsymbol{c}_1, \boldsymbol{c}_2, \ldots, \boldsymbol{c}_\ell) \in \mathcal{C}_A$.

---

**Algorithm 1: Decoding Procedure of Decoder $\mathcal{D}_A$**

**Input:** received word $\boldsymbol{y} = (\boldsymbol{y}_1, \boldsymbol{y}_2, \ldots, \boldsymbol{y}_\ell)$.
**Output:** codeword $\boldsymbol{c} \in \mathcal{C}_A$ or a decoding failure indicator "e".

1. Let $\boldsymbol{s}_j^1 = \boldsymbol{0}$, for $j = 1, 2, \ldots, \ell$.
2. $\hat{\boldsymbol{c}} = (\hat{\boldsymbol{c}}_1, \ldots, \hat{\boldsymbol{c}}_\ell) = \left(\mathcal{D}_1'(\boldsymbol{y}_1, \boldsymbol{s}_1^1), \ldots, \mathcal{D}_1'(\boldsymbol{y}_\ell, \boldsymbol{s}_\ell^1)\right)$.
3. Let $\mathcal{F} = \{j \in [\ell] : \hat{\boldsymbol{c}}_j \text{ contains } ?\}$.
4. **For** $i = 2, \ldots, \mu$
   - If $\mathcal{F} \neq \emptyset$, do the following steps; otherwise go to step 5.
   - $(\boldsymbol{s}_1^i, \cdots, \boldsymbol{s}_\ell^i) = \mathcal{D}_i''\left(\hat{\boldsymbol{c}}_1 H_i'^T, \ldots, \hat{\boldsymbol{c}}_\ell H_i'^T\right)$.
   - $\hat{\boldsymbol{c}}_j = \mathcal{D}_i'\left(\hat{\boldsymbol{c}}_j, (\boldsymbol{s}_j^1, \ldots, \boldsymbol{s}_j^i)\right)$ for $j \in \mathcal{F}$; $\hat{\boldsymbol{c}}_j$ remains the same for $j \in [\ell] \backslash \mathcal{F}$.
   - Update $\mathcal{F} = \{j \in [\ell] : \hat{\boldsymbol{c}}_j \text{ contains } ?\}$.
   **end**
5. If $\mathcal{F} = \emptyset$, let $\boldsymbol{c} = \hat{\boldsymbol{c}}$ and output $\boldsymbol{c}$; otherwise return "e".

---

In Algorithm 1, we use the following rules for operations which involve the symbol ?: 1) Addition $+$: for any element $\gamma \in \mathbb{F}_q \cup \{?\}$, $\gamma + ? = ?$. 2) Multiplication $\times$: for any element $\gamma \in \mathbb{F}_q \cup \{?\} \backslash \{0\}$, $\gamma \times ? = ?$, and $0 \times ? = 0$. 3) If a length-$n$ vector $\boldsymbol{x}$, $\boldsymbol{x} \in (\mathbb{F}_q \cup \{?\})^n$, contains an entry ?, then $\boldsymbol{x}$ is considered as the symbol ? in the set $\mathbb{F}_{q^n} \cup \{?\}$. Similarly, the symbol ? in the set $\mathbb{F}_{q^n} \cup \{?\}$ is treated as a length-$n$ vector whose entries are all ?s.

**Theorem 6.** *The decoder $\mathcal{D}_A$ for a $(\rho, n_0, k; d_0, d)_q$ ME-LRC $\mathcal{C}_A$ can correct any $d-1$ erasures in the array, and additional rows, each of which has less than or equal to $d_0 - 1$ erasures.*

*Proof:* See Appendix C. ■

## IV. Several Families of ME-LRCs Over Small Fields

In this section, we study the optimality of Construction A, and also present several explicit ME-LRCs that are optimal over different fields.

### A. Optimal Construction

We show how to construct ME-LRCs which are optimal with respect to the bound (1) by adding more constraints to Construction A. To this end, we specify the choice of the matrices in Construction A. This specification, referred to as **Design I**, is as follows.

1) $H_1'$ is the parity-check matrix of an $[n', n' - v_1, d_1']_q$ code which satisfies $k_{opt}^{(q)}[n', d_1'] = n' - v_1$.

2) $B_\mu$ is the parity-check matrix of an $[n', n' - \sum_{i=1}^\mu v_i, d_\mu']_q$ code with $d_{opt}^{(q)}[n', n' - \sum_{i=1}^\mu v_i] = d_\mu'$.

3) The minimum distances satisfy $d_\mu' \leqslant 2d_1'$.

4) $H_i''$ is an all-one vector of length $\ell$ over $\mathbb{F}_{q^{v_i}}$, i.e., the parity-check matrix of a parity code with minimum distance $\delta_i = 2$, for all $i = 2, \ldots, \mu$. ■

**Theorem 7.** *The code $\mathcal{C}_A$ from Construction A with Design I is a $(\rho = \ell, n_0 = n', k = n'\ell - v_1\ell - \sum_{i=2}^\mu v_i; d_0 = d_1', d = d_\mu')_q$ ME-LRC, which is optimal with respect to the bound* (1).

*Proof:* From Theorem 5, the code parameters $\rho$, $n_0$, $k$, $d_0$, and $d$ can be determined. We have $k^* = k_{opt}^{(q)}[n', d_1'] = n' - v_1$. Setting $x = \ell - 1$, we get

$$d \leqslant \min_{0 \leqslant x \leqslant \lceil \frac{k}{k^*} \rceil - 1} \left\{ d_{opt}^{(q)}[\rho n_0 - xn_0, k - xk^*] \right\}$$
$$\leqslant d_{opt}^{(q)}[\ell n' - (\ell - 1)n', k - (\ell - 1)k^*]$$
$$= d_{opt}^{(q)}[n', n' - \sum_{i=1}^\mu v_i] = d_\mu'.$$

This proves that $\mathcal{C}_A$ achieves the bound (1). ■

### B. Explicit ME-LRCs from Construction A

We give several explicit ME-LRCs over different fields. From Construction A, the local codes should have a nested structure and large minimum distance. Reed-Solomon codes and BCH codes are such codes. More generally, some cyclic codes and algebraic geometry codes [21] also have these properties. In the following, some explicit ME-LRCs are constructed over $\mathbb{F}_2$, $\mathbb{F}_3$, and $\mathbb{F}_4$.

*1) ME-LRCs with local extended BCH codes over $\mathbb{F}_2$*

From the structure of BCH codes [17], there exists a chain of nested binary extended BCH codes: $\mathcal{C}_3 = [2^m, 2^m - 1 - 3m, 8]_2 \subset \mathcal{C}_2 = [2^m, 2^m - 1 - 2m, 6]_2 \subset \mathcal{C}_1 = [2^m, 2^m - 1 - m, 4]_2$.

Let the matrices $B_1$, $B_2$, and $B_3$ be the parity-check matrices of $\mathcal{C}_1$, $\mathcal{C}_2$, and $\mathcal{C}_3$, respectively.

**Example 2.** For $\mu = 3$, in Construction A, we use the above matrices $B_1$, $B_2$, and $B_3$. We also choose $H_2''$ and $H_3''$ to be the all-one vector of length $\ell$ over $\mathbb{F}_{2^m}$.

From Theorem 5, the corresponding $(\rho, n_0, k; d_0, d)_2$ ME-LRC $\mathcal{C}_A$ has parameters $\rho = \ell$, $n_0 = 2^m$, $k = 2^m\ell - (m+1)\ell - 2m$, $d_0 = 4$, and $d = 8$. This code satisfies the requirements of Design I. Thus, from Theorem 7, it is optimal with respect to the bound (1). □

*2) ME-LRCs with local cyclic codes over $\mathbb{F}_3$*

Let $q = 3$ and $n' = 13$. We have the following factorization of $x^{13} - 1$ over $\mathbb{F}_3$:

$$x^{13} - 1 = (x+2)(x^3 + 2x + 2)(x^3 + x^2 + 2)$$
$$(x^3 + x^2 + x + 2)(x^3 + 2x^2 + 2x + 2)$$

We choose the matrices $B_1$, $B_2$, and $B_3$ in Construction A as follows. Let $B_1$ be the parity-check matrix of a $[13, 10, 3]_3$ code $\mathcal{C}_1$ with the generator polynomial:

$$g_1 = x^3 + x^2 + x + 2,$$

$B_2$ be the parity-check matrix of a $[13,6,6]_3$ code $\mathcal{C}_2$ with the generator polynomial:

$$g_2 = (x+2)(x^3+x^2+2)(x^3+x^2+x+2),$$

and $B_3$ be the parity-check matrix of a $[13,3,9]_3$ code $\mathcal{C}_3$ with the generator polynomial:

$$g_3 = (x+2)(x^3+x^2+2)(x^3+x^2+x+2)$$
$$(x^3+2x^2+2x+2).$$

It is clear that $\mathcal{C}_3 \subset \mathcal{C}_2 \subset \mathcal{C}_1$.

**Example 3.** For $\mu = 2$, in Construction A, we use the above matrices $B_1$ and $B_2$. We also choose $H_2^{''}$ to be the all-one vector with length $\ell$ over $\mathbb{F}_{3^4}$. From Theorem 5, the corresponding $(\rho, n_0, k; d_0, d)_3$ ME-LRC $\mathcal{C}_A$ has parameters $\rho = \ell$, $n_0 = 13$, $k = 10\ell - 4$, $d_0 = 3$, and $d = 6$. From Theorem 7, this code is optimal.

For $\mu = 3$, in Construction A, we use the above matrices $B_1$, $B_2$, and $B_3$. We also choose $H_2^{''}$ to be the parity-check matrix of an $[\ell, \ell-2, 3]_{81}$ MDS code where $\ell \leqslant 81$, and choose $H_3^{''}$ to be the all-one vector with length $\ell$ over $\mathbb{F}_{3^3}$. From Theorem 5, the corresponding $(\rho, n_0, k; d_0, d)_3$ ME-LRC $\mathcal{C}_A$ has parameters $\rho = \ell$, $n_0 = 13$, $k = 10\ell - 11$, $d_0 = 3$, and $d = 9$. We can calculate upper and lower bounds on the minimum distance of this ME-LRC from Lemma 2 and Lemma 3 respectively. For example, for $3 \leqslant \rho \leqslant 5$, the upper bound on $d$ is 12 and the lower bound on $d$ is 5. □

*3) ME-LRCs with local algebraic geometry codes over $\mathbb{F}_4$*

Algebraic geometry codes usually have large minimum distance and often possess the nested structure [21]. We will use a class of algebraic geometry codes called Hermitian codes [23] as an example to construct ME-LRCs.

From the construction of Hermitian codes [23], there exists a chain of nested 4-ary Hermitian codes: $\mathcal{C}_H(1) = [8,1,8]_4 \subset \mathcal{C}_H(2) = [8,2,6]_4 \subset \mathcal{C}_H(3) = [8,3,5]_4 \subset \mathcal{C}_H(4) = [8,4,4]_4 \subset \mathcal{C}_H(5) = [8,5,3]_4 \subset \mathcal{C}_H(6) = [8,6,2]_4 \subset \mathcal{C}_H(7) = [8,7,2]_4$. Note that except codes $\mathcal{C}_H(1)$ and $\mathcal{C}_H(7)$, other five codes are almost MDS codes, i.e., $n-k+1-d = 1$.

Now, let the matrices $B_1$, $B_2$, $B_3$, and $B_4$ be the parity-check matrices of $\mathcal{C}_H(4)$, $\mathcal{C}_H(3)$, $\mathcal{C}_H(2)$, and $\mathcal{C}_H(1)$, respectively. Let $H_i^{''}$, $i = 2,3,4$, be the all-one vector with length $\ell$ over $\mathbb{F}_4$.

**Example 4.** For $\mu = 2$, in Construction A, we use the above matrices $B_1$, $B_2$, and $H_2^{''}$. From Theorem 5, the corresponding $(\rho, n_0, k; d_0, d)_4$ ME-LRC $\mathcal{C}_A$ has parameters $\rho = \ell$, $n_0 = 8$, $k = 4\ell - 1$, $d_0 = 4$, and $d = 5$.

For $\mu = 3$, in Construction A, we use the above matrices $B_1$, $B_2$, $B_3$, $H_2^{''}$, and $H_3^{''}$. From Theorem 5, the corresponding $(\rho, n_0, k; d_0, d)_4$ ME-LRC $\mathcal{C}_A$ has parameters $\rho = \ell$, $n_0 = 8$, $k = 4\ell - 2$, $d_0 = 4$, and $d = 6$.

For $\mu = 4$, in Construction A, we use the above matrices $B_i$, $i = 1, \ldots, 4$, and $H_j^{''}$, $j = 2,3,4$. From Theorem 5, the corresponding $(\rho, n_0, k; d_0, d)_4$ ME-LRC $\mathcal{C}_A$ has parameters $\rho = \ell$, $n_0 = 8$, $k = 4\ell - 3$, $d_0 = 4$, and $d = 8$.

All of the above three families of ME-LRCs over $\mathbb{F}_4$ are optimal with respect to the bound (1). □

## V. CONCLUSION

In this work, we studied bounds and constructions for ME-LRCs. We presented a general construction that yields optimal ME-LRCs with a wide range of parameters when $d \leqslant 2d_0$. A corresponding erasure decoder was also proposed and shown to correct any $d - 1$ erasures. Finally, several families of explicit ME-LRCs were constructed over different small fields.

## APPENDIX A
## PROOF OF LEMMA 3

*Proof:* We can construct a $(\rho, n_0, k; d_0, d)_q$ ME-LRC in two steps, and use the GV bound [17] twice. First, there exists a $[\rho(n_0 - r_0), k, d]$ array code $\mathcal{G}_1$ of size $\rho \times (n_0 - r_0)$ where $r_0$ is an integer $0 \leq r_0 < n_0$, if it satisfies

$$\sum_{i=0}^{d-2} \binom{\rho(n_0 - r_0) - 1}{i} (q-1)^i < q^{\rho(n_0 - r_0) - k}. \quad (5)$$

Second, there exists a length-$n_0$ code $\mathcal{G}_2$ with minimum distance $d_0$, if its redundancy $r_0$ satisfies

$$r_0 > \log_q \Big( \sum_{i=0}^{d_0 - 2} \binom{n_0 - 1}{i} (q-1)^i \Big). \quad (6)$$

Now, we encode each row of the code $\mathcal{G}_1$ using the code $\mathcal{G}_2$ by adding $r_0$ more redundancy symbols. The resulting code is a $(\rho, n_0, k; d_0, d)_q$ ME-LRC. Let $r_0 = \lceil \log_q \big( \sum_{i=0}^{d_0 - 2} \binom{n_0 - 1}{i} (q-1)^i \big) \rceil$, and substitute it into (5), producing (3). ∎

## APPENDIX B
## PROOF OF THEOREM 4

*Proof:* A codeword $x$ in $\mathcal{C}_{GTP}^\mu$ is an $n'\ell$-dimensional vector over $\mathbb{F}_q$, denoted by $x = (x_1, x_2, \ldots, x_\ell)$, where $x_i$ in $x$ is an $n'$-dimensional vector, for $i = 1, 2, \ldots, \ell$.

Let $s_i^j = x_i H_j'^T$, for $i = 1, 2, \ldots, \ell$ and $j = 1, 2, \ldots, \mu$. Thus, $s_i^j$ is a $v_j$-dimensional vector over $\mathbb{F}_q$, and is considered as an element in $\mathbb{F}_{q^{v_j}}$. Let $s^j = (s_1^j, s_2^j, \ldots, s_\ell^j)$, an $\ell$-dimensional vector over $\mathbb{F}_{q^{v_j}}$, whose components are $s_i^j$, $i = 1, 2, \ldots, \ell$.

To prove Theorem 4, we need to show that if $xH^T = 0$ and $w_q(x) < d_m = \min\{\delta_1, \delta_2 d_1', \delta_3 d_2', \cdots, \delta_\mu d_{\mu-1}', d_\mu'\}$, then $x$ must be the all-zero vector $0$.

We prove it by contradiction and induction. Assume that there exists a codeword $x$ such that $xH^T = 0$, $w_q(x) < d_m$, and $x \neq 0$.

We first state a proposition which will be used in the following proof.

**Proposition 8.** *If $xH^T = 0$ and $s^1 = s^2 = \cdots = s^j = 0$, then $w_q(x_i) \geq d_j'$ for $x_i \neq 0$, $i = 1, 2, \ldots, \ell$.*

*Proof:* The condition $s^1 = s^2 = \cdots = s^j = 0$ means that $x_i B_j^T = 0$ for $i = 1, 2, \ldots, \ell$; that is, $x_i$ is a codeword in a code defined by the parity-check matrix $B_j$, whose minimum distance is $d_j'$. Therefore, we have $w_q(x_i) \geq d_j'$ for $x_i \neq 0$, $i = 1, 2, \ldots, \ell$. ∎

Now, if $s^1 \neq 0$, then $w_q(x) \geq w_{q^{v_1}}(s^1) \geq \delta_1 \geq d_m$, which contradicts the assumption. Thus, we have $s^1 = 0$.

Then, consider the second level. If $s^2 \neq 0$, then $w_q(x) \overset{(a)}{\geq} w_{q^{v_2}}(s^2) d_1' \geq \delta_2 d_1' \geq d_m$, where step (a) is from Proposition 8. Thus, it contradicts the assumption, so we have $s^2 = 0$. By induction, we must have $s^1 = s^2 = \cdots = s^{\mu-1} = 0$.

For the last level, i.e., the $\mu$th level, if $s^\mu \neq 0$, then $w_q(x) \geq w_{q^{v_\mu}}(s^\mu) d_{\mu-1}' \geq \delta_\mu d_{\mu-1}' \geq d_m$, which contradicts our assumption. Now, if $s^1 = s^2 = \cdots = s^\mu = 0$, then $w_q(x) \geq d_\mu' \geq d_m$, which also contradicts our assumption.

Thus, our assumption is violated. ∎

## APPENDIX C
## PROOF OF THEOREM 6

*Proof:* The ME-LRC $\mathcal{C}_A$ has $d_0 = d_1'$ and $d = d_\mu'$. Assume there exist $d_\mu' - 1$ erasures and additional $\rho'$ rows, each with less than or equal to $d_1' - 1$ erasures. We show that the decoder $\mathcal{D}_A$ can correct these erasures. The proof proceeds by induction.

For the first level, since $\delta_1 = \infty$, the correct syndrome vector $(s_1^1, \cdots, s_\ell^1)$ is the all-zero vector, i.e., $(s_1^1, \cdots, s_\ell^1) = 0$. Thus, the $\rho'$ rows are corrected. The remaining uncorrected row (block) $\hat{c}_j$, $j \in \mathcal{F}$, has at least $d_1'$ erasures. The total number of such rows (blocks) is less than $\delta_2$, because if it is $\delta_2$ or larger, then we have $\delta_2 d_1' \leq d_\mu' - 1$, which violates $d_\mu' \leq \delta_2 d_1'$.

Thus, for the second level, the correct syndrome vector $(s_1^2, \cdots, s_\ell^2)$ can be obtained. Similarly, the left uncorrected row $\hat{c}_j$, $j \in \mathcal{F}$, has at least $d_2'$ erasures, and the total number of such rows is less than $\delta_3$.

By induction, if the decoder runs until the $\mu$th level, then all the correct syndrome vectors $(s_1^i, \cdots, s_\ell^i)$, $i = 1, 2, \ldots, \mu$, are known. There must be only one uncorrected row left with number of erasures at least $d_{\mu-1}'$ and at most $d_\mu' - 1$. Thus, it can be corrected. ∎